\documentstyle[11pt,newpasp,twoside,epsf]{article}
\def\sax{{\it BeppoSAX}}

\def\edcomment#1{\iffalse\marginpar{\raggedright\sl#1\/}\else\relax\fi}
\reversemarginpar

\begin{document}
\title{Six Years of Gamma Ray Burst Observations with \sax}
\author{Filippo Frontera}
\affil{Physics Department, University of Ferrara, Via Paradiso, 12, 44100 Ferrara, Italy and 
IASF, CNR, Via Gobetti, 101, 40129 Bologna, Italy}

\begin{abstract}
I give a summary of the prompt X--/gamma--ray detections of Gamma Ray Bursts (GRBs) with 
the \sax\ satellite  
and discuss some significant results obtained from the study of the prompt
emission of these GRBs obtained with the \sax\ Gamma Ray Burst Monitor and Wide Field Cameras.
\end{abstract}

\section{The BeppoSAX revolution}

All the efforts performed before \sax, an Italian satellite with Dutch 
participation (Boella et al. 1997a), of unveiling the mystery of  Gamma Ray Bursts (GRBs), 
had not solved the problem. The most relevant result in this direction was obtained with the 
BATSE experiment aboard the {\it Compton} Gamma Ray Observatory (CGRO): the GRBs
were isotropically but not homogeneously distributed in the sky. These facts were very
suggestive of an extragalactic origin, but they were not a direct demonstration. 
Indeed, it was recognized (e.g., Fishman \&
Meegan 1995) that only the identification of a GRB counterpart at other wavelengths  was 
needed in order to make a breakthrough in setting their distance scale. Moreover
the presence of a delayed emission was not considered a necessary consequence the
observed GRB phenomenon. Thus, about thirty years after the initial discovery, the GRBs
were still a mysterious phenomenon. 

Thanks to \sax, this situation was reversed. The distance scale issue of long 
($>1$~s) GRBs  has been definitely  settled, the GRB afterglow emission at multiwavelengths 
has been discovered, with the GRB phenomenon, which is being  unveiled. \sax\ has opened a new 
window in the GRB astronomy, providing most of the exciting results of the last 6 years.  
The high performance of \sax\ for GRB studies is due to a well--matched configuration 
of its payload, with both wide field instruments (WFIs) and narrow field telescopes (NFIs). 
The WFIs comprised a $\gamma$--ray (40--700 keV) all--sky monitor (Gamma-Ray Burst Monitor, 
GRBM, Frontera et al. 1997) and two Wide Field Cameras (WFCs, 2--28 keV, Jager et al. 1997).
The NFIs included four focusing X-ray (0.1--10 keV) telescopes (one LECS, Parmar et al. 1997;
and three MECS, Boella et al. 1997b) and two higher energy 
direct--viewing detectors (HPGSPC, Manzo et al. 1997, and PDS, Frontera et al. 1997). 

The \sax\ capability of precisely (arcmin accuracy) localizing GRBs  was soon tested, 
only one
month after the first light of the satellite, with the detection of
GRB960720 (Piro et al. 1998). Starting from December 1996, at the \sax\
Science Operation Center, an alert procedure for the prompt search of simultaneous
GRBM and WFC detections of GRBs was implemented. 
Thanks to this procedure, the first X-ray afterglow of a
GRB which occurred on February 28, 1997  was soon discovered (8 hrs delay) and
its fading law determined (Costa et al. 1997). The afterglow light curve, unlike all other
transients, decayed according to a power--law. Simultaneously, an optical counterpart of 
the X-ray afterglow source was detected, which showed the same decay rate (van 
Paradijs et al. 1997). A ROSAT observation of the NFI error box confirmed the decay law 
of the fading source and established its positional coincidence with the optical
transient (Frontera et al. 1998). 

Less than two months later, the first optical redshift ($z = 0.835$) of a GRB source 
was determined thanks to the
prompt localization (5 hrs delay), provided by \sax, of GRB970508 (Metzger et al. 1997).
The extragalactic distance scale of the first GRB was definitely
established. At the same time an exciting result came from the radio observations 
of the same event: the
discovered radio afterglow exhibited, for about one month, rapid 
variations interpreted as interstellar scintillations. Only a very 
compact source highly relativistically expanding could be responsible
for them (Frail et al. 1997). This result was the first actual confirmation 
of fireball model for GRBs (see, e.g., review by Piran 1999).

Since that first detection of an X-ray afterglow, many other afterglow
detections have been obtained, mainly with \sax\ and, less often, with
other satellites, i.e., {\it ROSAT}, {\it Rossi} X--ray Time
Explorer (RXTE), {\it ASCA}, {\it Chandra} X--ray Observatory (CXO),
{\it XMM-Newton}. In this paper I summarize the GRBs promptly localized with \sax\
and discuss some relevant results obtained using GRBM and WFCs.

\section{Six Years of BeppoSAX results}

Table~1 reports the main information on the 51 GRBs simultaneously detected with GRBM and 
WFCs. This information includes positioning accuracy obtained with the WFCs,
duration of the events in the 2--10 and 40--700 keV energy bands, peak fluxes in these
two bands and the time delay between the GRB onset and the first TOO observation
with the \sax\ NFIs.

\begin{table}{t}
\small
\caption{GRBs detected with the \sax\ GRBM and WFCs}
\begin{tabular}{ccccccc}
\hline
 GRB & Position & 2--10 keV &T$_{{\rm 2-10 keV}}$ & 40--700 keV &T$_{{\rm 40-700 keV}}$ & 1$^{st}$ TOO \\ 
   & error radius&  peak flux & (s) & peak flux & (s) & delay  \\
   &    (arcmin)     & ($^{\mathrm a}$) &  & ($^{\mathrm a}$) &  & (hrs) \\            
\hline\noalign{\smallskip}
GRB960720 & 3 & 0.25 & 17 & 10 & 8 & 1080   \\
GRB970111 & 3 & 1.4 & 60 & 56 & 43 & 16  \\ 
GRB970228 & 3 & 1.4 & 80 & 37 & 80 & 8 \\
GRB970402 & 3 & 0.16 & 150 & 3.2 & 150 & 8 \\ 
GRB970508 & 1.9 & 0.35 & 29 & 5.6 & 15 & 5.7  \\
GRB971214 & 3.3 & 0.2 & 35 & 6.8 & 35 & 6.7  \\
GRB971227 & 8 & 0.36 & 7 & 3.8 & 7 & 14 \\
GRB980109 & 10 & 0.16 & 20 & 3.9 & 20 & --  \\
GRB980326 & 8 & 0.84 & 9 & 2.7 & 9 & -- \\
GRB980329 & 3 & 1.3 & 68 & 51 & 58 & 7  \\
GRB980425 & 8 & 0.61 & 40 & 2.4  & 31 & 9 \\
GRB980515 & 4 & 0.3 & 20 & 3 & 15 & 10  \\
GRB980519 & 3 & 0.51 & 190 & 13 & 28 & 7  \\
GRB980613 & 4 & 0.13 & 50 & 1.6 & 50 & 9  \\ 
GRB981226 & 6 & 0.085 & 260 & 0.56 & 20 & 11   \\
GRB990123 & 2 & 0.5 & 100 & 170 & 100 & 6  \\ 
GRB990217 & 3 & 0.11 & 25 & 1.1 & 25 & 6  \\
GRB990510 & 3 & 1.4  & 80 & 24. & 75 & 8 \\
GRB990625 & 60 & 0.065 & 11  &  0.11 & 11 &  -- \\
GRB990627 & 3 & 0.065 & 60  &  1.6 & 28 &  8   \\
GRB990704 & 7 & 1.3  & 40 &  1.8 & 23 &  8  \\
GRB990705 & 3 & 0.85 & 45   &  37  & 42 &  11 \\
GRB990712 & 2 &  3.1 & 30 & 6.3  & 30 &  -- \\
GRB990806 & 3 & 0.35 & 30 & 5.5  & 30 &   8 \\
GRB990907 & 6 & 0.39  &  220  & 9.5   &  1  &  11  \\
GRB990908 & 8 & 0.35  & 130  & 1.0   & 50  &  -- \\ 
GRB991014 & 6 &   0.54  &  10  &  4.2 & 3   &  13  \\
GRB991105 & 5 &   0.24   &  40  &  6.4 & 13   &  --  \\
GRB000210 & 2 & 1.6  & 20 & 180  & 20  &  7  \\
GRB000214 & 6.5 & 0.63 & 115  & 40  & 10   &  12   \\
GRB000528 & 2 & 0.62 & 120  & 14  & 80   &  12  \\
GRB000529 & 4 & 0.2  &  30    & 7   & 14   &  7.5    \\
GRB000615 & 2 &  0.8 &  120  & 1   & 12   &  10    \\
GRB000620 & 3.6 &  0.9    &  20   & 7    & 15   &  --    \\
GRB001011 & 2 &  0.07  &  60  & 25    & 31   & --  \\
GRB001109 & 2.5 &  0.6    &  65   & 4.2    & 60   & 16.5   \\
GRB010213 & 6 &  0.65    &  25   & 14   & 23   & --  \\
GRB010214 & 3 &  1.7    &  30   & 5.5    & 20   & 6   \\
GRB010220 & 4 &  1.5    &  150   & 5.6    & 40   & 13.6    \\
GRB010222 & 2.5 &  2.1    &  280   & 86    & 170   & 8    \\
GRB010304 & 2.5 &  0.59    &  24    & 11    & 15   & 8    \\
GRB010412 & 1.6 &  0.64    &  90   & 17    & 74   & --   \\
GRB010501 & 6 &  0.09    &  41   & 1.5    & 37   & --    \\
GRB010518 & 5 &  0.15    &  30   & 1.3    & 25   & --   \\
GRB011121 & 2 &  3.65    &  100  & 73     & 105   & 21.2 \\
GRB011211 & 2 &  0.09    &  400  & 0.5    & 400  & 11.1  \\
GRB020321 & 5 &  0.06    &  90   & 1.1    & 70   & 8  \\
GRB020322 & 3 &  0.17    &  50   & 3.0    & 15   & 7.5  \\
GRB020409 & 3.2 &  0.36  &  60   & 1.8    & 40   & --   \\
GRB020410 & 2   & 0.19 &  $> 1290$  & 1.3 ($^{\mathrm b}$)  & 1800 ($^{\mathrm b}$) & 
     20.2 \\
GRB020427 & 3   & 0.19 &  60        & $<0.66$ (3$\sigma$) &   -      &  11.2  \\
\hline
\end{tabular}
($^{\mathrm a}$) Peak fluxes in units of 10$^{-7}$erg cm$^{-2}$ s$^{-1}$ \\
($^{\mathrm b}$) 15--1000 keV KONUS data. GRBM was switched off. 
\end{table}

As can be seen, all the GRB events detected with both WFCs and GRBM are long (>2 s), even 
if a sizable fraction (22
GRBM are short (Guidorzi et al. 2003a). In addition, most of the detected events are 
classical GRBs, except two, which are X-ray rich events (981226, 990704).
Four of the detected GRBs
(960720, 010213,010304, 010501)  were discovered with the off-line analysis, three
(010518, 020321, 020322) were alerted by the ground trigger software, 36 were followed up 
with the NFIs.

X--ray afterglows, whose measurements were performed starting from $\le 5.7$~hrs after the 
main event, are discovered in about 90\% of the followed-on GRBs, against  about 
50\%  in the optical and about 40\% in the radio. Thus an open problem is the origin
of  GRBs with no optical emission (dark). Are they very weak? Do not they emit  
optical radiation? Is this radiation strongly absorbed by the circumburst medium?
Are they at very large distances? Do they fade very rapidly? Likely some or more than 
one of the above reasons could be at the origin of the dark GRBs. With the future missions
(e.g., SWIFT), the origin of the dark GRBs will be  certainly unveiled.

\section {Some significant results}

I review below few of the most relevant results obtained with the GRBM plus WFC
instruments aboard \sax, in particular the discovery of X--ray rich GRBs,
the test of syncrotron shock model from the shape of the X--/gamma--ray prompt emission
spectra and their evolution, the discovered relation between peak
energy of the $\nu F(\nu)$ spectrum and the isotropic electromagnetic energy released in the
GRB prompt emission, the discovery of a transient absorption feature during
the rise time of the burst profile of GRB990705 (evidence of a transient absorption
feature from GRB011211 is under investigation, Frontera et al. 2003).

\subsection{X--ray rich GRBs: the link between classical GRBs and XRFs}

A new class of GRBs has been discovered with \sax. It includes GRBs which, 
unlike classical GRBs, emit most of their energy in the X--ray band (2--28 keV). 
Two oustanding examples of X--ray rich GRBs were detected with \sax: GRB981226 
(Frontera et al. 2000a), and GRB990704 (Feroci et al. 2001). Their light curves are 
shown in Fig.~1. As can be seen, both GRBs are well visible in the 2--26 keV band, 
but they are barely visible in  the
40--700 keV interval. Aslo a percursor with onset 180~s before the main event,
is only observed in the X--ray band, but not in gamma--rays. Gamma--ray emission is
only visible for a short time in correspondence of the GRB onset in the case of GRB981226,
and in correspondence of the main peak in the case of GRB990504.

A useful parameter which characterize the X--ray richness is the ratio between the
the fluences in the 2--10 keV and 40--700 keV energy band, respectively (softness ratio). 
Figure~2, 
from the paper by Feroci et al. (2001), shows the  behaviour of the softness ratio  for the
two X--ray rich GRBs compared with that of some classical GRBs. Their separation
from the classical GRBs is apparent. The softness ratio of X--ray flashes (XRF, 
Heise et al. 2001, not plotted
in the figure)  is still more separated, 
making X--ray rich events GRBs as the link between classical GRBs and XRFs. 

%
%
\begin{figure}
\plottwo{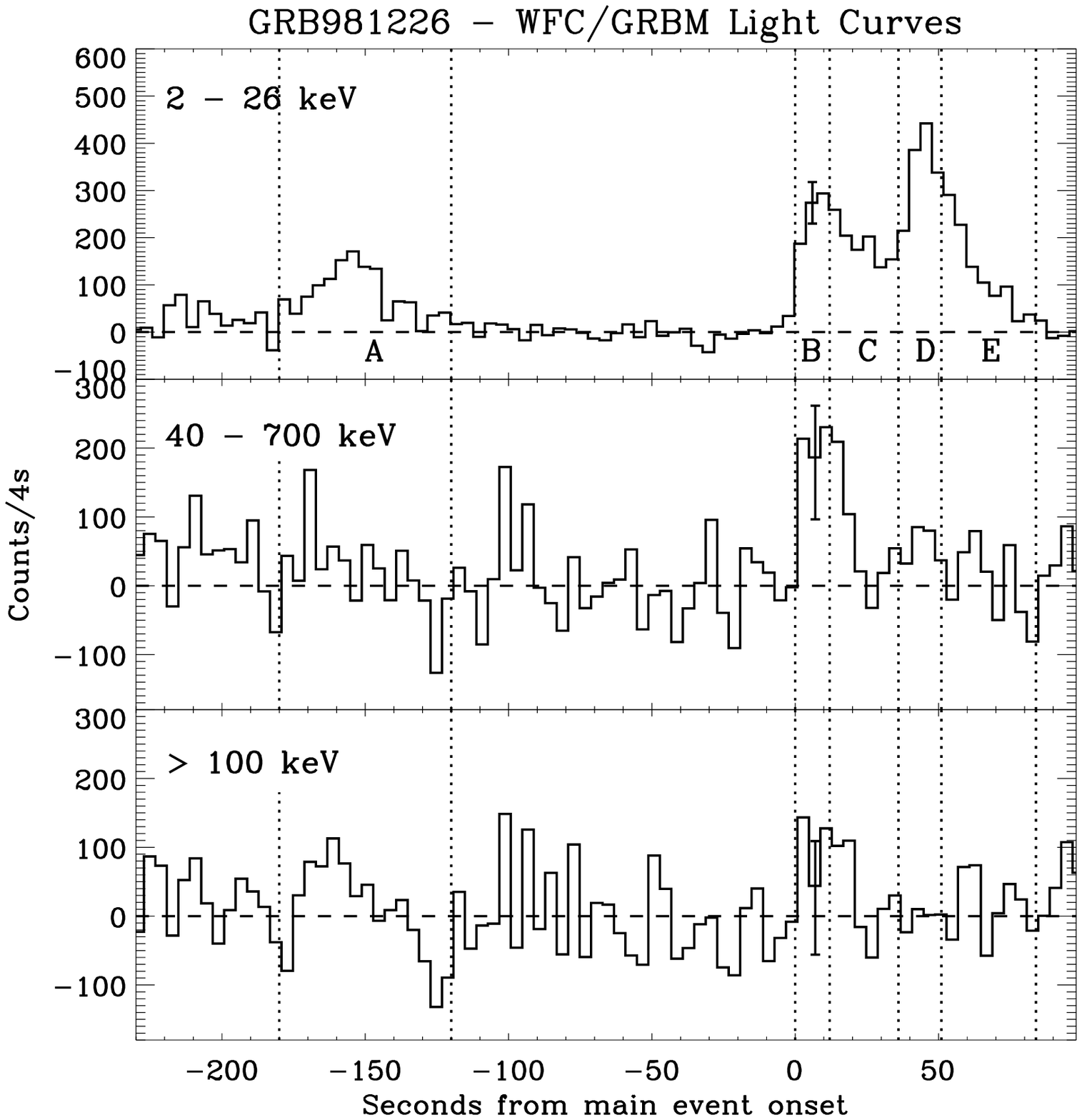}{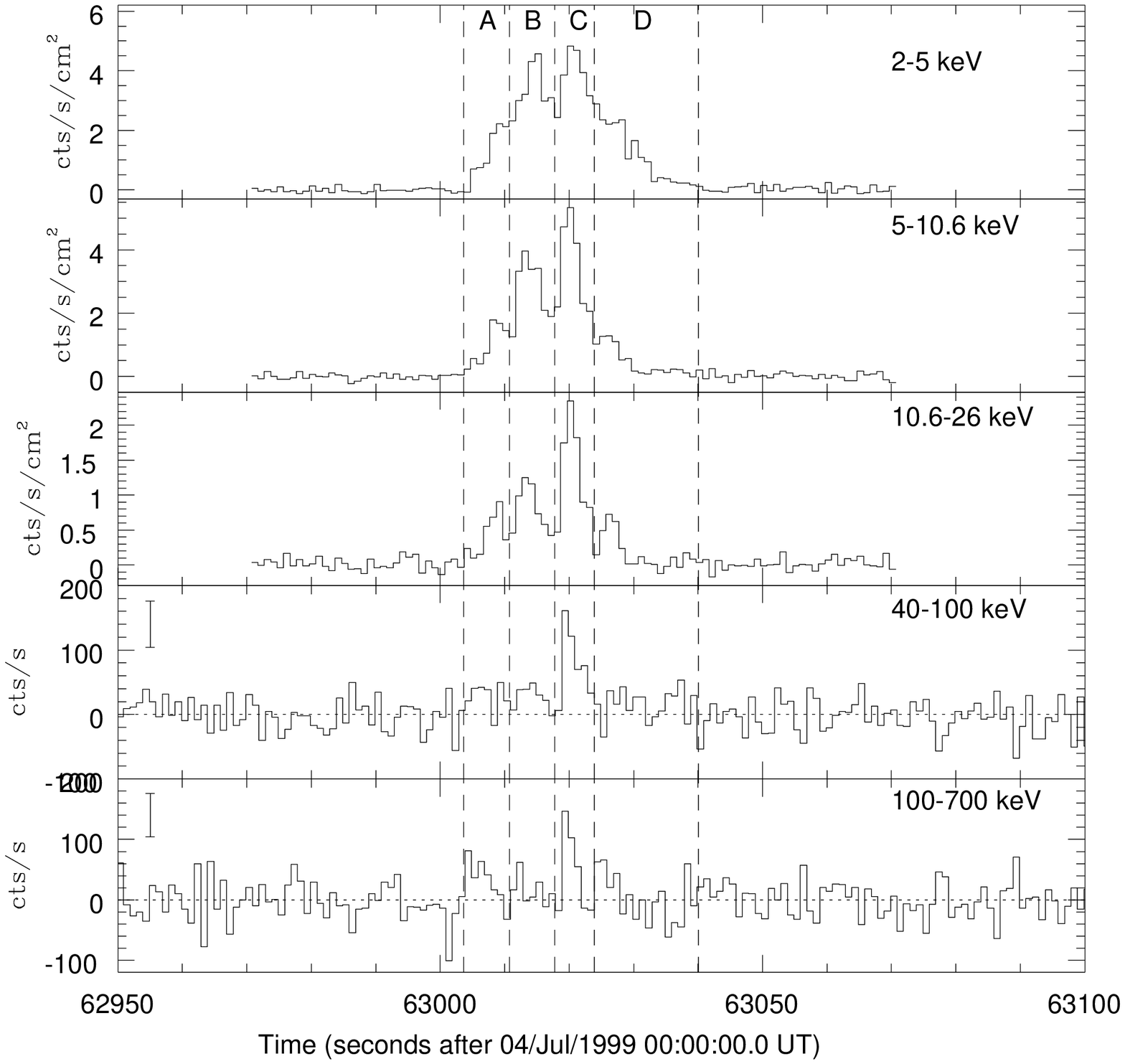}
\vspace{0.7cm}
\caption{Light curves of two X--ray rich GRBs. {\it Left}: GRB981226, reprinted from 
Frontera et al. (2000a). {\it Right}: GRB990704, reprinted from Feroci et al. (2001).}
\label{f:981226_990704_lc}
\end{figure}

%
%
\begin{figure}
\plotfiddle{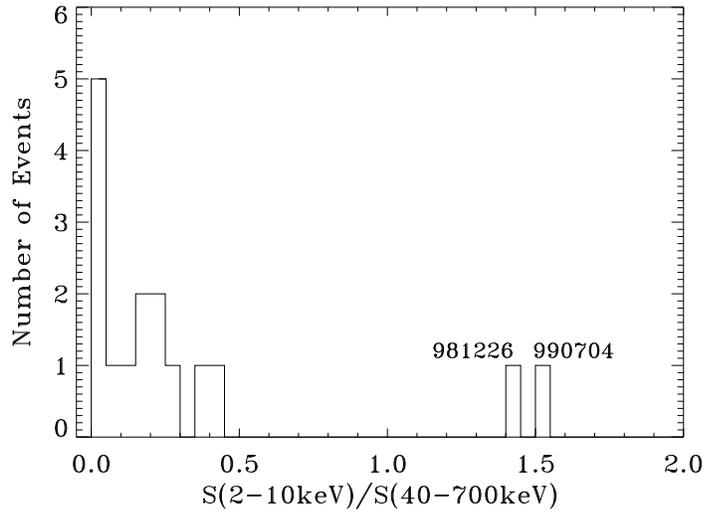}{7cm}{0}{60}{60}{-210}{-250}
\vspace{1.0cm}
\caption{Ratio $S$(2--10 keV)/$S$(40--700 keV) between X--ray and gamma--ray fluences for
various GRBs detected with the \sax\ WFC and GRBM. The position in the diagram
of the X--ray rich GRBs 981226 and 990704 is marked. Reprinted from Feroci et al. (2001).}
\label{f:fluence_ratio}
\end{figure}

Both two X--ray rich GRBs were followed-on with the \sax\ NFIs. 
The afterglow spectra are consistent with power--laws (photon indices $\Gamma$ of 
$1.92\pm 0.47$ for GRB981226 and $1.69^{+0.60}_{-0.34}$  forGRB990704), as 
those of many  classical GRBs.
Also the afterglow light curves shows decay rates typical of classical GRBs. 
However an unexpected dip of at least two hours in the afterglow decay rate
is observed at the beginning of the GRB981226 afterglow observation, which was started 11 
hrs after the main event. This feature was interpreted as due to a temporary cessation 
(or strong reduction) of the X--ray afterglow due to the absence (or very reduced density) 
of ambient gas, like a cavity surrounding the explosion (Frontera et al. 2000a).

\subsection{Test of the 2--700 keV prompt emission mechanism}

As discussed in various papers (e.g., Amati et al. 2002), most of the time averaged 
spectra of the GRBs observed with \sax\ WFCs and GRBM are well fit down to 2 keV with 
a smoothly broken power-law ('Band function', Band et al. 1993) (in the other cases, a 
single power--law describes the data). The Band function, as discussed by Tavani (1996),
has a spectral shape similar to that of the synchroton radiation if the energy spectrum 
of the emitting electrons is hybrid, i.e., it is distributed partially according to
a Maxwellian function and partially according to a power--law. The  synchrotron spectrum 
derived by Tavani (1996) also assumes a negligible self absorption
(Optically Thin Synchrotron Shock Model, OTSSM). We fit the time averaged spectra of 19
GRBs with this model (Amati et al. 2001). The result is that in about 70\% of the cases 
the OTSSM fit well the data. Two examples of these fit results are shown in 
Fig.~3. A property of the OTSS model is that, below
the peak energy of the $E F(E)$ spectrum, the photon index is $-2/3$, independently of
the electron energy distribution and uniformity of the associated magnetic field, if
no electron cooling takes place during the GRB. We find that this does not occur
 in 30\% of our data. In these cases, the spectral index is consistent with that expected
in the case of an electron cooling: photon index between $-2/3$ and $-1.5$).

%
%
\begin{figure}
\plottwo{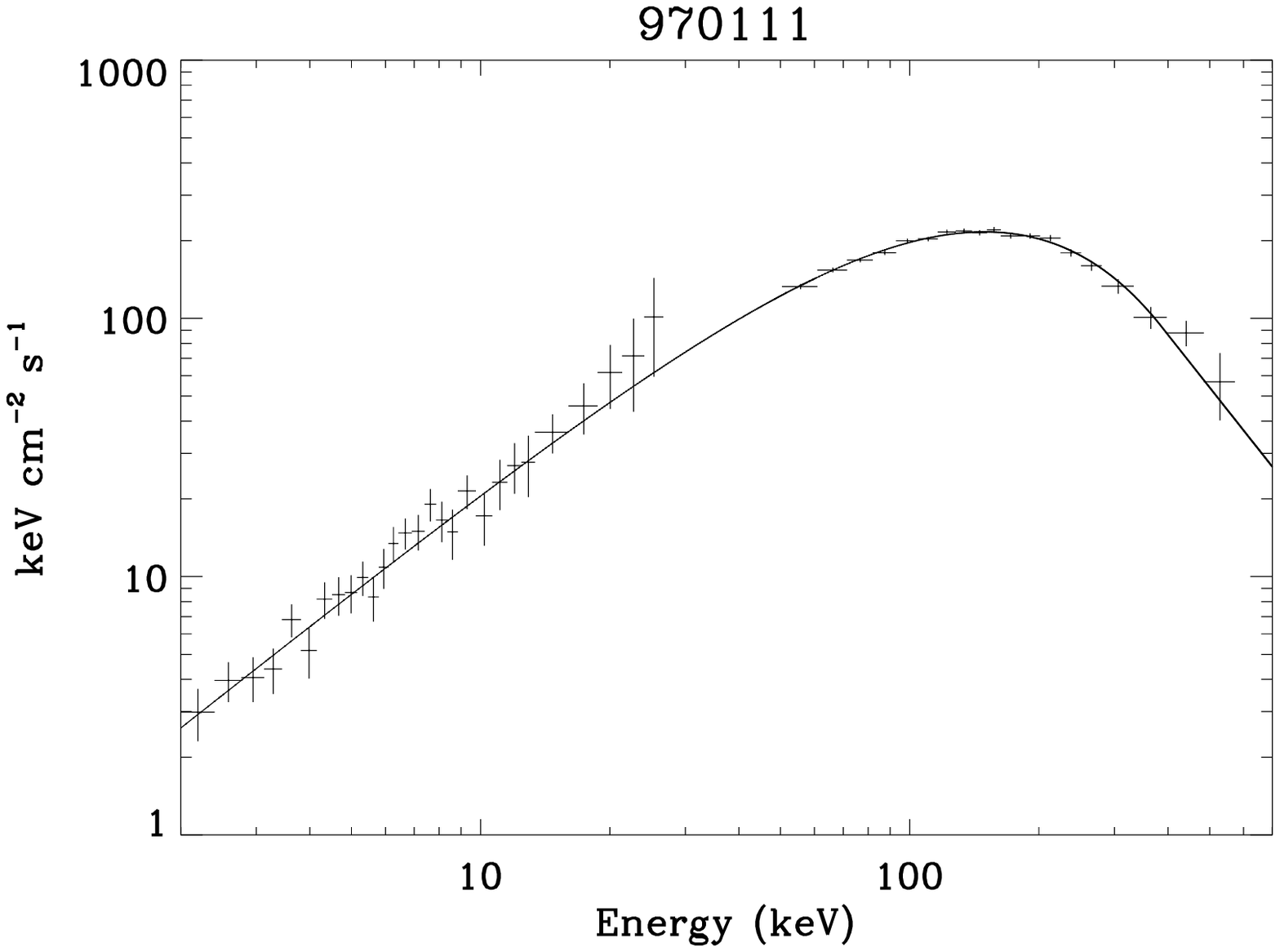}{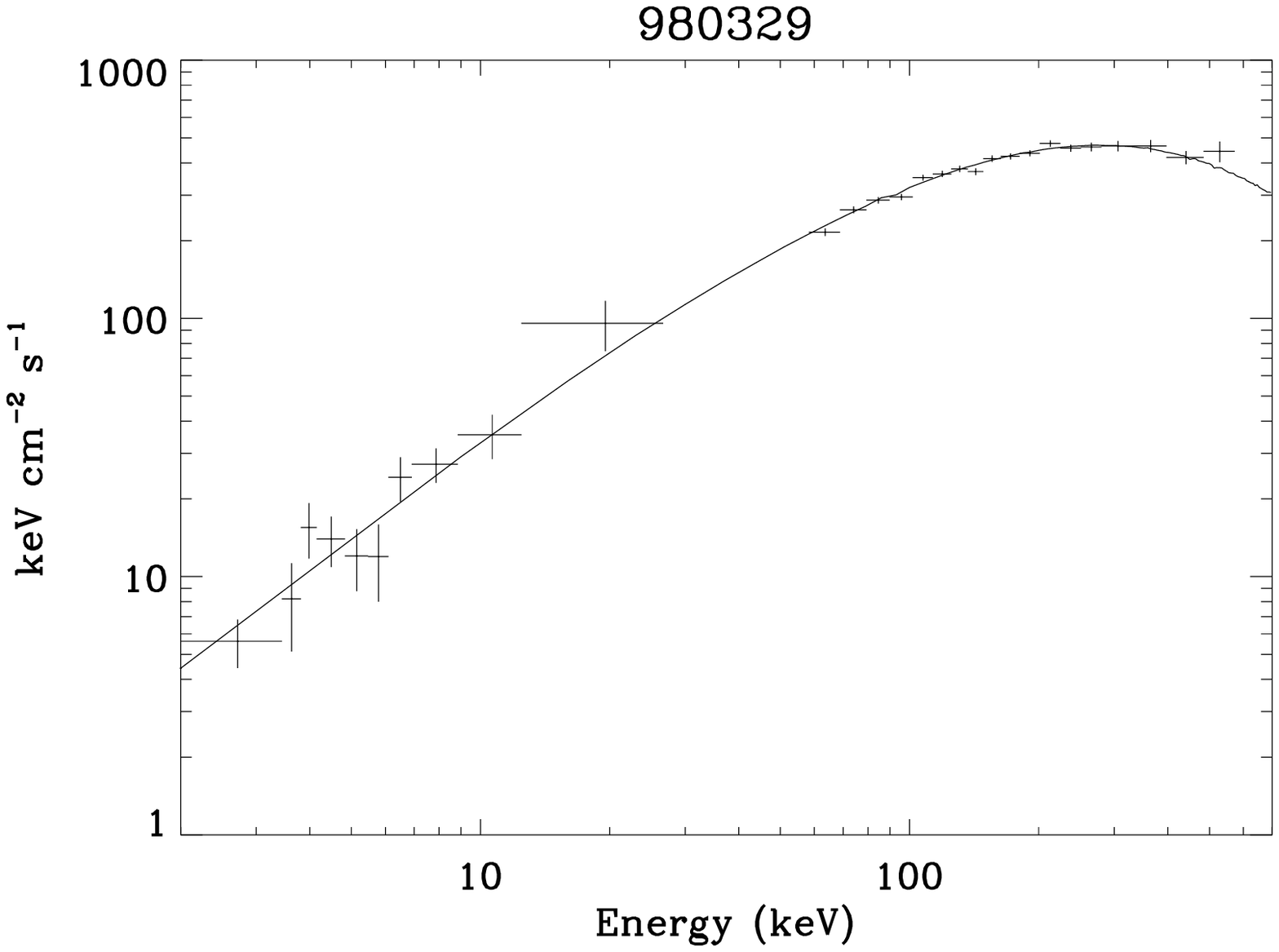}
\caption{Time averaged $E F(E)$ spectrum of GRB970111 ({\it left}) and GRB980329 ({\it right}), 
with superposed the fit with the optically thin synchrotron shock model (Tavani 1996).
Reprinted from Amati et al. 2001.}
\label{f:970111_980329_sp}
\end{figure}

A more constraining test of the prompt emission mechanism is the spectral evolution
of the GRB  emission. This study has been performed for almost all GRBs detected
with the \sax\ GRBM and WFCs (see Frontera et al. (2000b) for a sample of GRBs, and
Frontera et al. 2003a for the entire population of  \sax\ GRBM and WFCs events).
The analysis, in addition to a general hard to soft evolution, shows that in most cases 
the spectra soon after the GRB onset cannot be fit with an OTSSM, which however becomes 
acceptable as the time elapses. This feature shows that likely some other emission mechanism 
(e.g. Inverse Compton, as discussed by Frontera et al. 2000b) is at work at  the early times.

\subsection{Peak energy vs. isotropic released energy correlation}

An investigation devoted to search out correlations between parameters derived
from the redshift--corrected energy spectra of GRBs with known redshift has permitted
us to discover (Amati et al. 2002) a power--law relation (see Fig.~4)
between 
peak energy $E_p$ of the $E F(E)$ redshift--corrected time averaged spectra and
isotropic electromagnetic energy $E_{rad}$ released in the GRB event:
\begin{equation}
E_p \propto E_{rad}^{0.52\pm 0.06} 
\end{equation}
The relation is now confirmed (D. Lamb, private communication) by HETE results.
It puts strong constraints to the GRB emission models: independently of
the beaming, eq. (1) must be satisfied.
A discussion on  possible interpretations of the above relation is
given  by Zhang and M\'esz\'aros (2002) in the context of the internal and external shock
models.

%
%
\begin{figure}
\plotfiddle{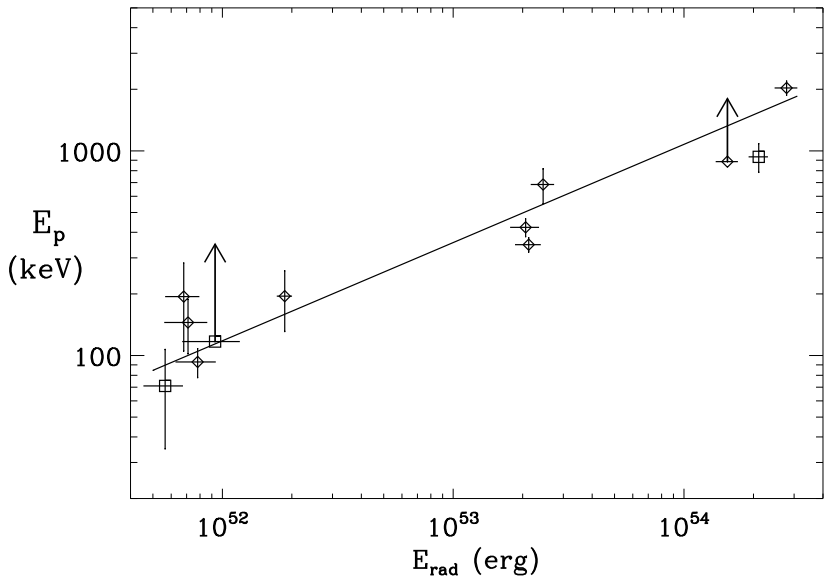}{7cm}{0}{67}{67}{-230}{-250}
\caption{Peak energy $E_p$ of the redshift-corrected energy $E F(E)$ spectra of
GRBs with known redshift as a function of the isotropic gamma--ray energy released during
prompt emission. Reprinted from Amati et al. 2002.}
\label{f:970111_sp_ev}
\end{figure}

\subsection{Test of the GRB environment}

With the WFC and GRBM spectral data, it has also been possibile to gain information
on the GRB environment from the study of the hydrogen-equivalent column 
density behaviour with time, and the search
of X--ray lines in the prompt emission. Both these investigations have given positive
results, with the discovery of variable hydrogen column density from some GRBs
and the discovery of a transient absorption line from at least one GRB.
We discuss both.

\subsubsection{Variable column density}

A decreasing $N_{\rm H}$  has been detected in the X--ray spectra of at least events 
three GRB events: GRB980329 (Frontera et al. 2000), GRB010222 (in 't Zand et al. 2001), 
GRB010214 (Guidorzi et al. 2003). 
The $N_{\rm H}$ time behaviour observed from GRB980329 is explained 
(Lazzati \& Perna 2001) 
if the GRB event occurs in an overdense region within a molecular cloud, with properties 
similar to those of star formation globules (Bok globules).

\subsubsection{Transient absorption feature from GRB990705}

A clear evidence of a transient absorption feature was found in the X--ray spectrum
collected during the rise (first 13 s) of GRB990705 (event total duration of about 40~s).

Two  possible interpretations have been proposed. Amati et al. (2000) interpreted the feature
as a K absorption edge of neutral Fe (rest frame energy $E_K = 7.1$~keV) within a shell of 
material around the GRB location, which is photo-ionized by the GRB photons. 
With this assumption, the  GRB redshift derived is $z_{X-rays} = 0.86\pm 0.17$, and the iron 
relative abundance is $A/A_\odot \sim 75$, which is typical of a young supernova explosion
environment.
A successful test of this model is the later measurement of the redshift ($z_{opt} = 
0.8420 \pm 0.0002$ of the GRB host galaxy (Le Floc'h et al. 2002), which is nicely
consistent with the X--ray redshift value. The drawback of this model is the
large mass of Fe implied (several solar masses), unless the Fe material is clumped 
and a clump was along the line of sight (Boettcher et al. 2001).
 In order to overcome this problem, Lazzati et al. (2001) interpreted the absorption
feature as an absorption line due resonant scattering of GRB photons off H--like Fe 
(transition 1s-2p, $E_{rest} = 6.927$~keV). With this assumption, the X--ray redshift is
still consistent with that of the GRB host galaxy, the line broadening is interpreted
as dispersion in the outflow velocity (up to ~0.1c), the required relative Fe abundance 
($A/A_\odot \sim 10$) is still consistent with the site of a young supernova explosion, but
the Fe mass required is only $\sim 0.2$ M$_\odot$.

Thus, independently of the specific model, the observed feature points to an iron-rich 
circumburst environment, in which a supernova explodes first, and the GRB occurs later on.
These results are consistent with the expectations of the 'supranova' model for GRB
progenitors (Vietri \& Stella 1998).

Evedence of another X-ray transient absorption line from GRB011211 is under evaluation
(Frontera et al. 2003b).

\section{Conclusions}

\sax\ has given a key contribution  to the study  of the GRB prompt 
emission properties in an unprecedented broad energy band (2-700 keV).
Certainly many questions are left unsolved by \sax, which hopefully will 
be answered by the future missions (e.g., SWIFT), but I am certain that the 
six years of BeppoSAX will be unlikely left down.

\acknowledgments

I wish to thank Lorenzo Amati for his help to prepare Table 1 of this paper.
I also acknoledge the support by the Italian Spage Agency ASI and the Ministry of
Education, University and Research of Italy (COFIN funds 2001).


\begin{references}

Amati, L. et al. 2000, {\it Science}, 290, 953

Amati, L. et al. 2001, in: {\it Gamma--Ray Bursts in the Afterglow Era},
  eds. E. Costa, F. Frontera \& J. Hjorth (Springer, Berlin), p. 34

Amati, L. et al. 2002, \aap, 390, 81

Band, D. et al. 1993, \apj, 413, 281

Boella, G. et al. 1997a, \aaps, 122, 299

Boella, G. et al. 1997b, \aaps, 122, 327

Boettcher, M. et al. 2001, {\it Gamma--Ray Bursts in the Afterglow Era},
  eds. E. Costa, F. Frontera \& J. Hjorth (Springer, Berlin), p. 160

Costa, E. et al. 1997, {\it Nature}, 387, 783

Feroci, M. et al. 2001, \aap, 378, 441

Fishman, G.A. \& Meegan, C.A. 1995, \araa, 33, 415 (1995)

Frail, D.A. et al. 1997, {\it Nature}, 389, 261

Frontera, F. et al. 1997, \aaps, 122, 357

Frontera, F. et al. 1998, \aap, 334, L69 

Frontera, F. et al. 2000a, \apj, 540, 697

Frontera, F. et al. 2000b, \apjs, 127, 59

Frontera, F. et al. 2003a, in preparation

Frontera, F. et al. 2003b, in preparation

Guidorzi, C. et al. 2003a, these Proc.s

Guidorzi, C. et al. 2003b, \aap, 401, 491

Heise, J. et al. 2001, in: {\it Gamma--Ray Bursts in the Afterglow Era},
  eds. E. Costa, F. Frontera \& J. Hjorth (Springer, Berlin), p. 16

Jager, R. et al. 1997, \aaps, 125, 557

Lazzati, D. \& Perna, R. 2001, \mnras, 330, 383

Lazzati et al. 2001, \apj, 556, 471

Le Floc'h, E. et al. 2002, \apj, 581, L81

Manzo, G. et al. 1997, \aaps, 122, 341

Metzger, M.R. et al. 1997, {\it Nature}, 387, 878

Parmar, A.N. et al. 1997, \aaps, 122, 309 

Piran, T. 1999, {\it Phys. Rep.},  314, 575

Piro, L. et al. 1998, \aap, 329, 906

Tavani, M. 1996, \apj, 466, 768

Vietri, M. \& Stella, L. 1998, \apj, 507, L45

van Paradijs, J. et al. 1997, {\it Nature}, 386, 686 

in 't Zand, J.J.M. et al. 2001, \apj, 559, 710
 
Zhang, B. \& M\'esz\'aros, P. 2002, \apj, 581, 1236

\end{references}
\end{document}